\documentclass[iop,apj]{emulateapj}
\usepackage{hyperref}
\usepackage{amsmath,amstext}
\usepackage[T1]{fontenc}
\usepackage{apjfonts} 
\usepackage[figure,figure*]{hypcap}
\usepackage{graphicx}
\usepackage{multirow}

\def\f28{${f}_{2-8{\rm keV}}$}

\def\ergs{erg s$^{-1}$}
\def\ergscm2{erg s$^{-1}$ cm$^{-2}$}
\def\yr-1{yr$^{-1}$}

\def\asec{\ifmmode^{\prime\prime}\else$^{\prime\prime}$\fi}

\def\secspt{$\buildrel{\prime\prime}\over .$}

\def\Chandra{{\it Chandra}}

\def\Swift{{\it Swift}}

\shorttitle{Deep Chandra Study of GW170817} 
\shortauthors{Haggard {\it et al.}}

\begin{document}

\title{A Deep {\it Chandra} X-ray Study of Neutron Star Coalescence GW170817}

\author{
Daryl Haggard$^{1,2}$, Melania Nynka$^{1}$, John J.~Ruan$^{1}$, Vicky Kalogera$^{3}$, S. Bradley Cenko$^{4,5}$, Phil Evans$^{6}$, Jamie A. Kennea$^{7}$
}
\affil{
$^{1}$McGill Space Institute and Department of Physics, McGill University, 3600 rue University, Montreal, Quebec, H3A 2T8, Canada \\
$^{2}$CIFAR Azrieli Global Scholar, Gravity \& the Extreme Universe Program, Canadian Institute for Advanced Research, 661 University Ave., Suite 505, Toronto, Ontario M5G 1M1, Canada \\
$^{3}$Center for Interdisciplinary Exploration and Research in Astrophysics and Department of Physics and Astronomy, Northwestern University, 2145 Sheridan Road, Evanston, Illinois 60208-3112, USA \\
$^{4}$Astrophysics Science Division, NASA Goddard Space Flight Center, Greenbelt, MD 20771, USA \\
$^{5}$Joint Space-Science Institute, University of Maryland, College Park, MD 20742, USA \\
$^{6}$Leicester Institute for Space and Earth Observation and Department of Physics \& Astronomy, University of Leicester, University Road, Leicester, LE1 7RH, UK \\
$^{7}$Department of Astronomy and Astrophysics, Pennsylvania State University, 525 Davey Lab, University Park, PA 16802, USA
}

\begin{abstract}
We report \Chandra\ observations of GW170817, the first neutron star--neutron star merger discovered by the joint LIGO-Virgo Collaboration, and the first direct detection of gravitational radiation associated with an electromagnetic counterpart, {\it Fermi} short $\gamma$-ray burst GRB 170817A. The event occurred on 2017 August 17 and subsequent observations identified an optical counterpart, SSS17a, coincident with NGC 4993 ($\sim$10\asec\ separation). Early \Chandra\ ($\Delta t\sim2$ days) and \Swift\ ($\Delta t \sim1-3$ days) observations yielded non-detections at the optical position, but $\sim$9 days post-trigger \Chandra\ monitoring revealed an X-ray point source coincident with SSS17a. We present two deep \Chandra\ observations totaling $\sim$95 ks, collected on 2017 September 01--02 ($\Delta t \sim 15-16$ days). We detect X-ray emission from SSS17a with $L_{0.3-10 {\rm keV}} = 2.6^{+0.5}_{-0.4}\times10^{38}$ \ergs, and a power law spectrum of $\Gamma=2.4\pm0.8$. We find that the X-ray light curve from a binary NS coalescence associated with this source is consistent with the afterglow from an off-axis short $\gamma$-ray burst, with a jet angled $\gtrsim$23$^\circ$ from the line of sight. This event marks both the first electromagnetic counterpart to a LIGO-Virgo gravitational-wave source and the first identification of an off-axis short GRB. We also confirm extended X-ray emission from NGC 4993 ($L_{0.3-10 {\rm keV}} \sim 9\times10^{38}$ \ergs) consistent with its E/S0 galaxy classification, and report two new {\it Chandra} point sources in this field, CXOU J130948 and CXOU J130946.
\end{abstract}


\section{Introduction}
\label{sec:intro}

The detection of gravitational waves (GWs) by the Laser Interferometer Gravitational-wave Observatory (LIGO) is one of the most exciting advances in physics in decades. \citet{abbott16a} reported the first LIGO detection of GWs, resulting from the merger of two black holes (BHs). The observed waveforms showed a near-perfect match to predictions from general relativity for the inspiral and merger of two black holes, ushering in the era of gravitational-wave astronomy. Extensive follow-up observations based on this GW event found no robust electromagnetic (EM) counterparts \citep[e.g.,][]{abbott16b, connaughton16, evans16, soares16}, consistent with theoretical predictions for stellar-mass BH mergers.

The next frontier is multi-messenger astronomy, where GW sources are associated with an EM emitter, connecting GW astronomy to our rich understanding of astrophysics. Core-collapse supernovae, mergers of two neutron stars (NSs), and mergers of NS-BH binaries are among the EM sources likely to have detectable GW signals. In particular, NS-NS mergers have long been predicted to be the progenitors of short $\gamma$-ray bursts \citep[GRBs;][]{paczynski86,narayan92}, and may produce kilonovae \citep{li98} that are responsible for the majority of $r$-process nucleosynthesis in the Universe \citep{eichler89}.

On 17 August 2017 at 12:41:04 UTC, LIGO-Virgo detected event GW170817---its observed waveform traced the distinctive signal of a NS-NS inspiral and early analysis indicated a luminosity distance of $D_L = 40\pm8$ Mpc \citep{ligo21509}. This discovery is the first in a new class of GW events stemming from neutron star binary coalescences, which are predicted to produce EM emission. Approximately 2 seconds after the GW trigger, the Gamma-ray Burst Monitor (GBM) instrument onboard the \emph{Fermi} Gamma-ray Space Telescope was also triggered by the short-duration GRB 170817A \citep{connaughton21506,vonkienlin21520,goldstein21528, goldstein17}. Thanks to tight localization by LIGO-Virgo, follow-up ground-based optical imaging soon discovered the associated optical transient Swope Supernova Survey 17a \citep[SSS17a, ][]{coulter21529,coulter17}, near the galaxy NGC 4993 at $z=0.0098$ \citep[$D_L = 42.5\pm0.3$ Mpc;][]{daCosta98}. 

\begin{figure*} [t!]
\center{
\includegraphics[trim={3cm 0 4.3cm 0},clip,scale=0.50,angle=270]{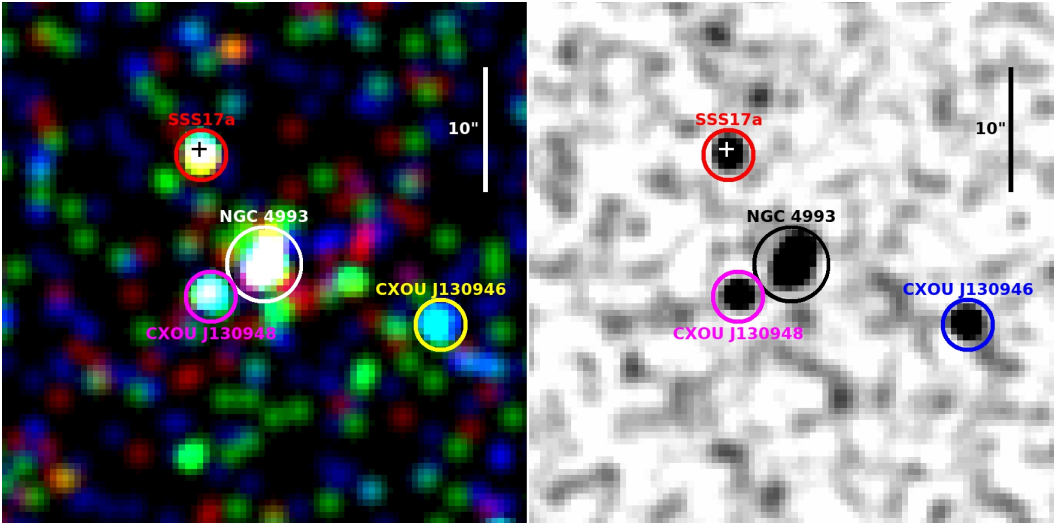}
}
\figcaption{\textit{\textbf{Left}}: Merged three-color X-ray image from \Chandra\ ObsIDs 20899 and 18988 covering a 42\asec$\times$42\asec\ field of view. The color channels represent energy ranges with red$~=0.5-1.2$ keV (soft), green$~=1.2-2.0$ keV (medium), and blue$~=2.0-7.0$ keV (hard). \textit{\textbf{Right}}: The same \Chandra\ patch showing the full $0.5-7.0$ keV energy range for the merged observations. Each panel is centered on the host galaxy NGC 4993 and contains four X-ray sources: SSS17a (red extraction circle), CXOU J130948 (magenta circle), CXOU J130946 (yellow [left] or blue [right] circle), and NGC 4993 (white [left] or black [right] circle). These have radii of 1\secspt97 for the point sources ($\sim$90\% encircled energy), and $2$\secspt$95$ for NGC~4993. The optical position of SSS17a \citep[R.A. $=13:09:48.085\pm0.060$, decl. $=-23:22:53.343\pm0.731$][]{coulter21529,coulter17} is shown as a cross, and a 10\asec\ bar is shown for scale. \Chandra's excellent $\sim0.5$\asec\ spatial resolution is crucial for identifying the X-ray counterpart and disentangling the flux from these individual X-ray sources.}
\label{fig:image}
\end{figure*}
 
This discovery initiated rapid follow-up surveillance by X-ray telescopes. The first X-ray observations of the field yielded upper limits from the Monitor of All-sky X-ray Images (MAXI) on board the International Space Station \citep{sugita21555} and the X-ray Telescope (XRT) on the \Swift\ Observatory \citep{evans21550}. In particular, \Swift\ observations began 0.6 days post-trigger, followed by a cadence of one-to-several observations daily. No X-ray emission was detected at the location of SSS17a down to a limiting luminosity of $L_{0.3-10~\mathrm{keV}}=9.2\times10^{38}$~erg~s$^{-1}$ \citep{evans17}. Stacked Swift-XRT observations spanning 16 days post-trigger revealed a possible weak source reported in \citet{evans21612} that, with refined astrometric corrections and additional exposure, was localized to R.A.=13:09:47.65, decl.=-23:23:01.6 with a $90\%$ confidence radius of 3\secspt9 \citep{evans17}. The \Swift\ position and luminosity, $L_{0.3-10~\mathrm{keV}}\sim4\times10^{39}$~erg~s$^{-1}$, are consistent with the host NGC 4993, but due to the $\sim 15$\asec\ point spread function of Swift, it is possible that the nearby X-ray point source CXOU J130948 contaminates both (Figure \ref{fig:image}). 

Prior to the observations reported here, \Chandra\ also observed the field of NGC~4993. The first observation occurred approximately 2 days post-trigger and reported a non-detection at the location of SSS17a \citep{margutti21648,margutti17}. An observation 9 days post-trigger detected a source consistent with SSS17a, though no flux or luminosity values were reported \citep{troja21765,troja17}. 

In this Letter, we present two deep \Chandra\ X-ray observations of the field of GW170817. In a 42\asec$\times$42\asec\ patch centered on NGC 4993 we detect four X-ray sources, including SSS17a and spatially-extended X-ray emission from the host galaxy. By constructing a \Chandra\ X-ray light curve of SSS17a using these and earlier \Chandra\ observations, we show that the X-ray emission from this NS-NS merger is consistent with the afterglow from an off-axis short GRB, with a jet axis angle of $\gtrsim$23$^\circ$. If confirmed, this makes GRB 170817A the first off-axis short GRB observed to date, in addition to being the first EM counterpart to a LIGO-Virgo GW detection.

\begin{figure}[h!]
\center{
\includegraphics[scale=0.45]{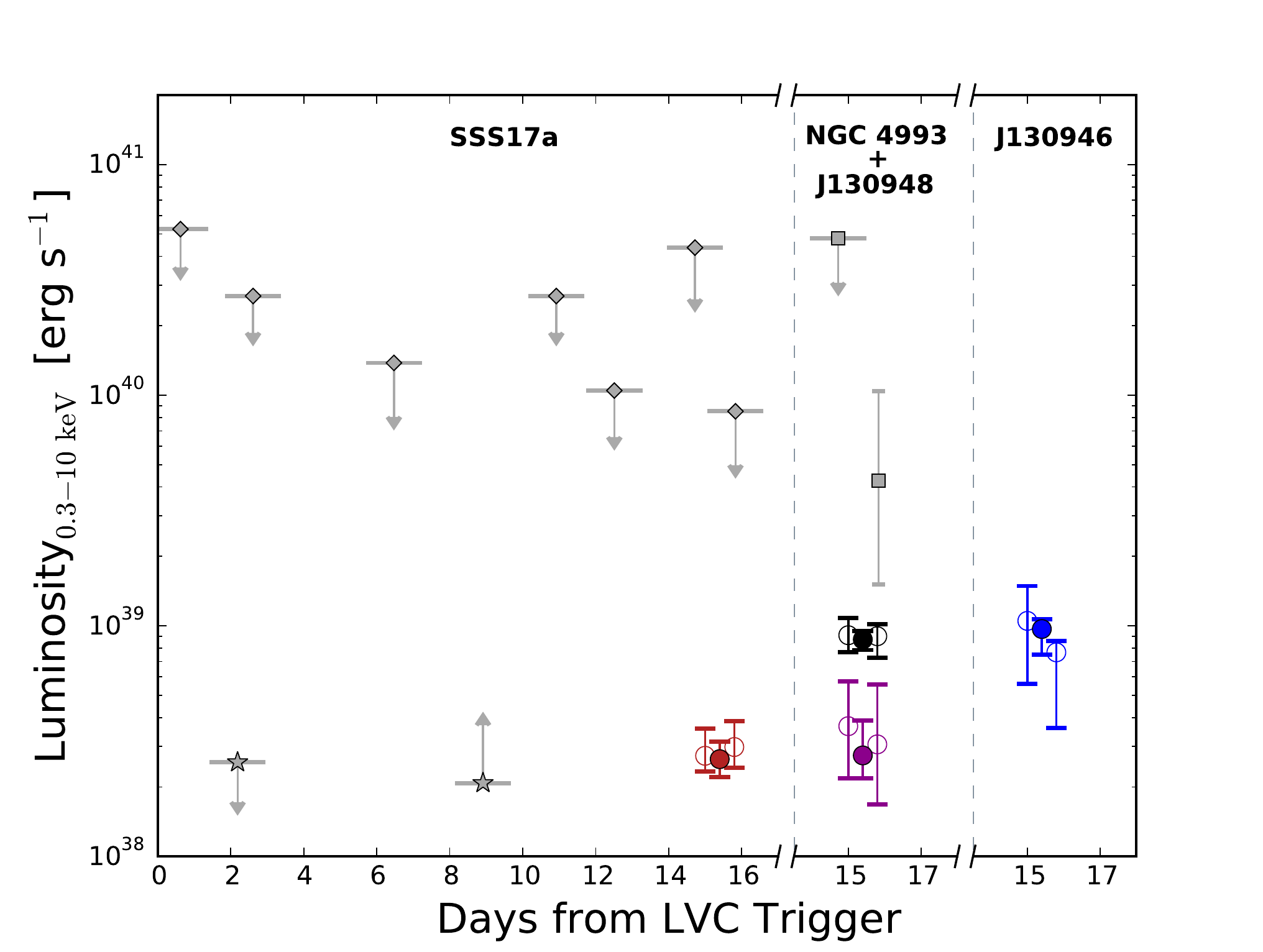}
}
\figcaption{ \Chandra\ X-ray luminosities for individual sources, marked as colored symbols: SSS17a (red), NGC 4993 (black), CXOU J130948 (magenta), and CXOU J130946 (blue). Empty circles represent values obtained from individual spectra analyzed here (ObsIDs 20899 and 18988); filled circles are luminosities from co-added spectra. For SSS17a, grey stars indicate \Chandra\ upper \citep{margutti21648} and lower \citep{troja21765} limits and grey diamonds mark \Swift\ upper limits \citep{evans21550}. \Swift\ luminosities for NGC 4993 (which are likely blended with CXOU J130948) are shown as grey squares \citep{evans17}.}
\label{fig:fluxes}
\end{figure}

The outline of this Letter is as follows: In Section \ref{sec:obssec}, we describe our \Chandra\ observations. In Section \ref{sec:xray_prop}, we discuss the properties of the X-ray sources in the field of GW170817. In Section \ref{sec:discussion}, we interpret our results and summarize our conclusions. Throughout this Letter, we adopt a standard $\Lambda$CDM cosmology with $\Omega_\mathrm{M} = 0.31$, $\Omega_\Lambda = 0.69$, and $H_0 = 68$ km s$^{-1}$ Mpc$^{-1}$, consistent with the results of \citet{planck16}.

\section{Observations}
\label{sec:obssec}

We report analysis of two 46.69 ks \Chandra\ X-ray observations, ObsID 20899 and ObsID 18988, which cover a patch of the LIGO-Virgo high confidence localization for GW170817 (LIGO Scientific Collaboration and Virgo Collaboration GNCs 21505 \& 21509). ObsID 20899 (PI Troja) began 2017 September 01 at 15:22:22 ($\sim$15 days post-trigger) and ObsID 18988 (PI Haggard) began approximately 13 hours later on 2017 September 02 at 04:53:25 ($\sim$16 days post-trigger). Both observations were acquired using \Chandra's ACIS-S3 chip in VFAINT mode. Data reduction and analysis were performed with CIAO v.4.8 tools \citep[CALDB v4.7.2;][]{fruscione06}. We reprocessed the level 2 events files, applied the latest calibrations via CIAO's {\tt repro} script, and extracted the 0.5--7 keV images and X-ray spectra described in \S\ref{sec:xray_prop}. Our small field of view (Figure \ref{fig:image}) includes the optical transient SSS17a \citep{coulter21529,coulter17}, the Swift X-ray detection \citep{evans21550,evans21612}, and several other X-ray sources of interest.

Continued monitoring observations of this field with \Chandra\ (as well as \Swift) were prohibited by Sun constraints beginning in mid-September 2017 and continuing until early December 2017 (grey region in the left panel of Figure \ref{fig:lc}). 

\begin{deluxetable*}{lccccccc}
\tablecaption{X-ray Source Properties}
\tablehead{  
  \colhead{Source} & \colhead{RA} & \colhead{Dec} & \colhead{Extraction} & \colhead{{\it Chandra}} & \colhead{Power Law~$^a$} & \colhead{Flux (${0.3-8}$ keV)} & 
            \colhead{Luminosity (${0.3-10}$ keV)~$^b$} \\
  \colhead{ID} & \colhead{(J2000)} & \colhead{(J2000)} & \colhead{Radius} & \colhead{ObsID} & \colhead{$\Gamma$} & \colhead{[$10^{-14}~$erg~s$^{-1}$~cm$^{-2}$]} & \colhead{[$10^{38}$~erg~s$^{-1}$]}
  }
\startdata
SSS17a  & 13:09:48.077 & -23:22:53.459 & $1.968''$ & 20899 & \multirow{2}{*}{$2.7^{+1.0}_{-0.8}$} & $0.35^{+0.1}_{-0.04}$ & $2.7^{+0.8}_{-0.4}$ \vspace{3pt}\\
        &              &               &           & 18988 &                                      & $0.38^{+0.2}_{-0.04}$ & $3.0^{+0.9}_{-0.6}$ \vspace{3pt}\\
        &              &               &        & combined &  $2.4\pm0.8$                         & $0.36^{+0.1}_{-0.07}$ & $2.6^{+0.5}_{-0.4}$ \vspace{3pt}\\
\hline
\rule{0pt}{2.5ex}
CXOU J130948 & 13:09:48.014 & -23:23:04.917 & $1.968''$ & 20899 & \multirow{2}{*}{$1.0^{+0.9}_{-1.0}$} & $0.5^{+0.3}_{-0.2}$   & $3.7^{+2.1}_{-1.5}$ \vspace{3pt}\\
        &              &               &           & 18988 &                                      & $0.4\pm0.2$           & $3.1^{+2.5}_{-1.4}$ \vspace{3pt}\\
        &              &               &        & combined &  $1.3\pm0.8$                         & $0.4^{+0.1}_{-0.09}$  & $2.7^{+1.1}_{-0.6}$ \vspace{3pt}\\ 
\hline
\rule{0pt}{2.5ex}
CXOU J130946 & 13:09:46.682 & -23:22:06.983 & $1.968''$ & 20899 & \multirow{2}{*}{$-0.1^{+0.4}_{-0.9}$} & $1.2^{+0.4}_{-0.5}$ & $10.5^{+4.4}_{-4.9}$ \vspace{3pt}\\
        &              &               &           & 18988 &                                       & $0.9^{+0.3}_{-0.4}$ & $7.7^{+0.9}_{-4.1}$ \vspace{3pt}\\
        &              &               &        & combined &  $-0.4^{+0.2}_{-0.8}$                 & $1.1\pm0.1$         & $9.7^{+1.0}_{-2.2}$ \vspace{3pt}\\
\hline
\rule{0pt}{2.5ex}
NGC 4993 & 13:09:47.705 & -23:23:02.457 & $2.95''$  & 20899 & \multirow{2}{*}{$1.37^{+0.5}_{-0.4}$} & $1.4\pm0.2$         & $9.1^{+1.7}_{-1.4}$ \vspace{3pt}\\
         &              &               &           & 18988 &                                       & $1.3^{+0.3}_{-0.2}$ & $9.0^{+1.2}_{-1.7}$ \vspace{3pt}\\
         &              &               &        & combined &  $1.5\pm0.4$                          & $1.3\pm0.2$         & $8.7^{+0.8}_{-0.9}$
\enddata
\tablenotetext{}{
$^a$ The neutral hydrogen absorption was frozen to N\textsubscript{H}$=7.5 \times10^{20}$~cm$^{-2}$ for all spectral fits, based on NGC 4993's A$_{\rm V}=0.338$ \citep[][see Section \ref{sec:xray_prop} for details]{Schlafly11}.
$^b$ A luminosity distance of 42.5~Mpc was assumed for all sources. 
}
\label{tab:specvals}
\end{deluxetable*}

\section{X-ray Analysis and Source Properties}
\label{sec:xray_prop}

In a small, on-axis patch $\sim0.5'$ on a side (Figure \ref{fig:image}), we detected X-ray emission from three point sources and from one extended source: (1) point-source X-ray emission at the location of the optical transient SSS17a \citep{coulter21529,troja21765,troja21787,fong21786,haggard21798}, (2) another point source, CXOU J130948, near the location of the Swift X-ray emission \citep{evans21612,fong21786,haggard21798}, (3) emission from another previously unidentified X-ray point source, CXOU 130946, and (4) extended emission from the host galaxy NGC 4993 \citep{evans21612,margutti21648}.
 
We used CIAO's {\tt wavdetect} to obtain the centroid position for each source in the broadband ($0.5-7$~keV) images. We selected a 1\secspt97 extraction region for the three point sources, corresponding to a $\sim$90\% encircled energy fraction near \Chandra's aim point. A $2$\secspt$95$ extraction radius for NGC~4993 was chosen to enclose as much of its emission as possible while minimizing contamination from nearby CXOU~J130948. Care was also taken to extract background photons from a large region that did not enclose other sources. Source IDs, positions, and extraction regions are reported in Table \ref{tab:specvals} and visualized in Figure \ref{fig:image}.

\begin{figure*}[t!]
\center{
\includegraphics[scale=0.27,angle=90]{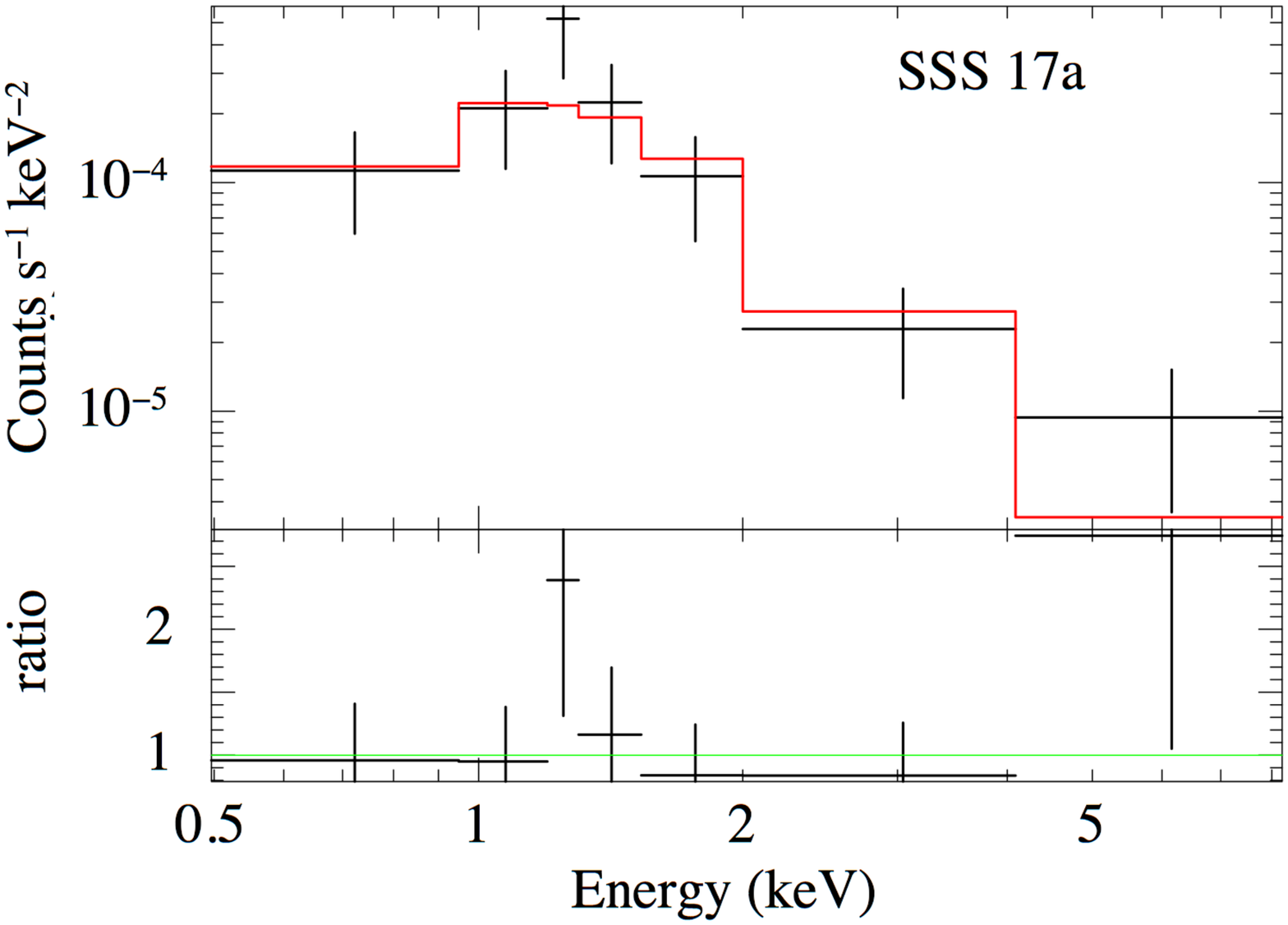}
\includegraphics[scale=0.27,angle=90]{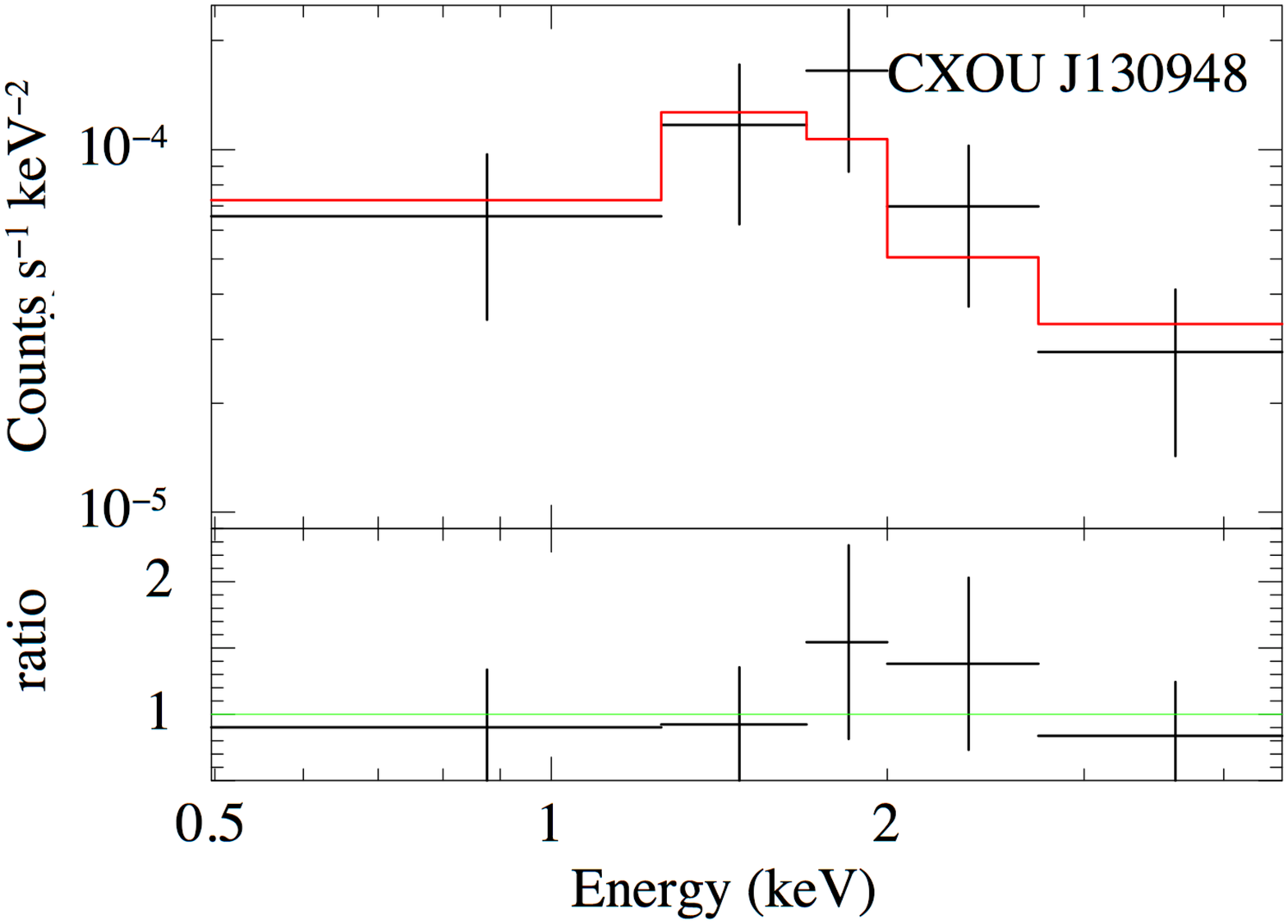}
\includegraphics[scale=0.28,angle=90]{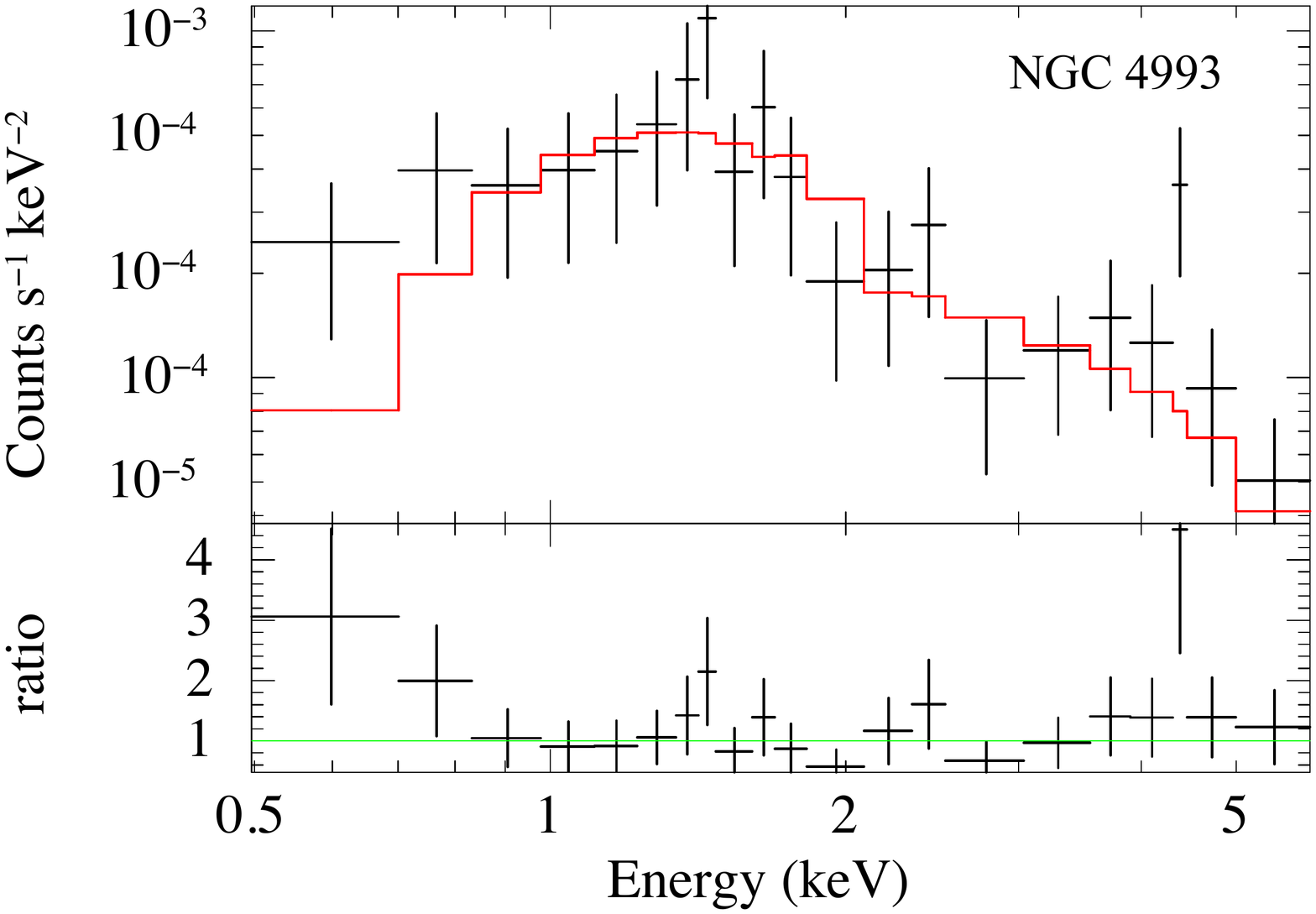}
\includegraphics[scale=0.27,angle=90]{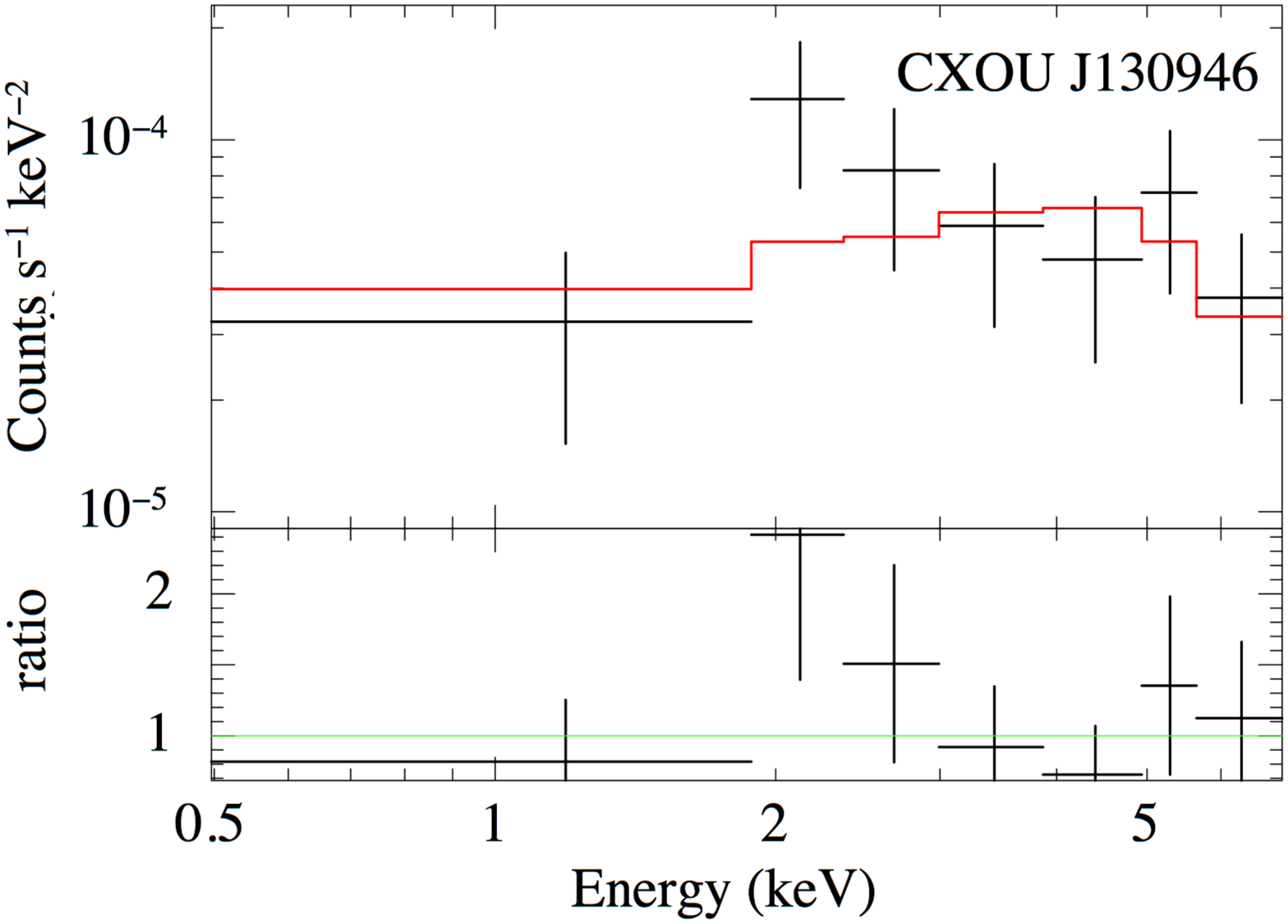}
}
\figcaption{ \Chandra\ co-added X-ray spectra from SSS17a (\textit{\textbf{Top Left}}), CXOU J130948 (\textit{\textbf{Top Right}}), NGC~4993 (\textit{\textbf{Bottom Left}}) and CXOU J130946 (\textit{\textbf{Bottom Right}}). 
Spectral data and residuals are shown in black and the best-fit spectral model is in red. 
The neutral hydrogen absorption column was fixed to 
N\textsubscript{H}$=7.5 \times10^{20}$~cm$^{-2}$ (see Section \ref{sec:xray_prop}).  For all jointly fit data the spectral index $\Gamma$ was tied between the observations while the power law normalization was left free. }
\label{fig:spec}
\end{figure*}

\begin{figure*}[t!]
\center{
\includegraphics[scale=0.62]{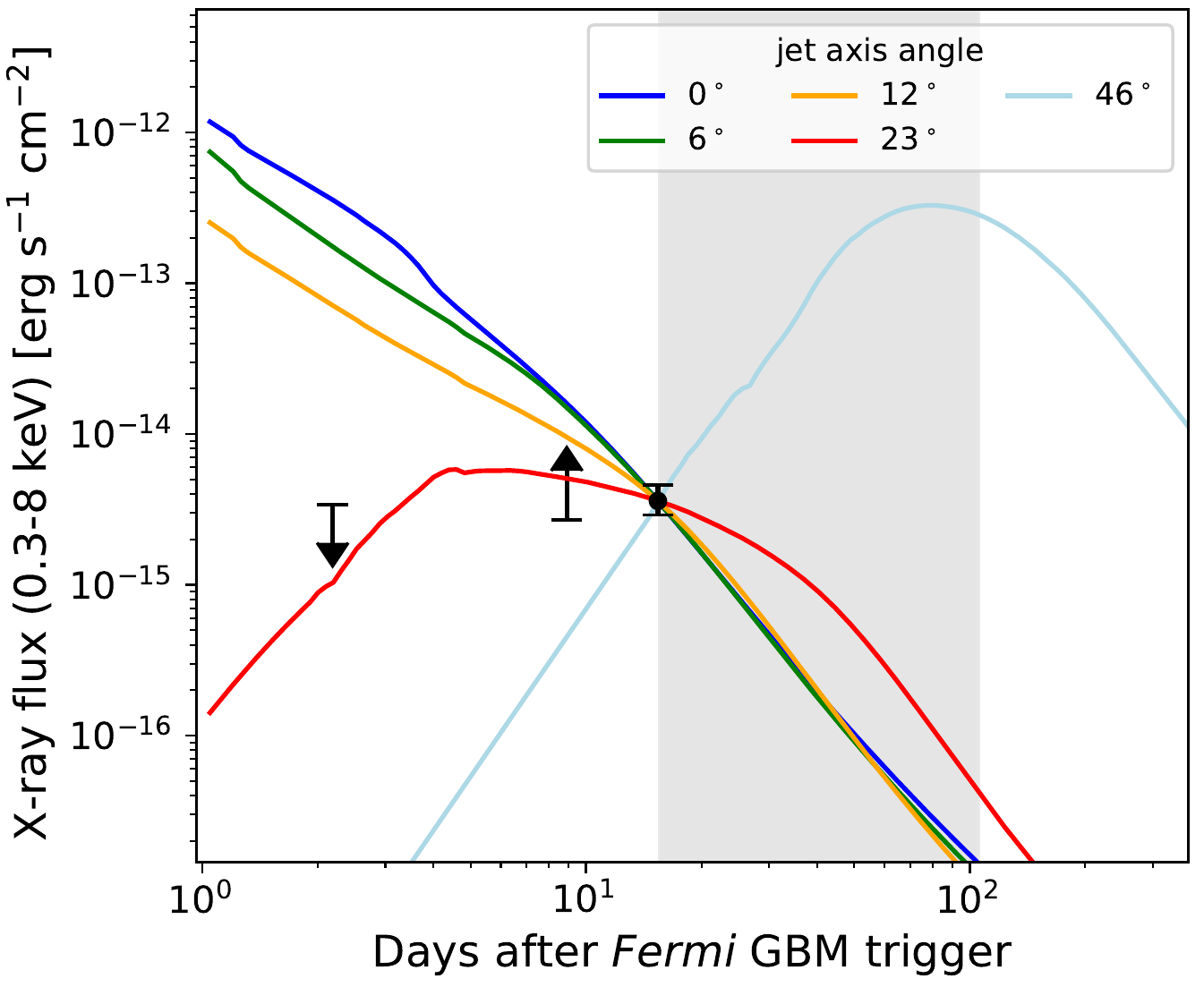}
\hspace{5pt}
\includegraphics[scale=0.62]{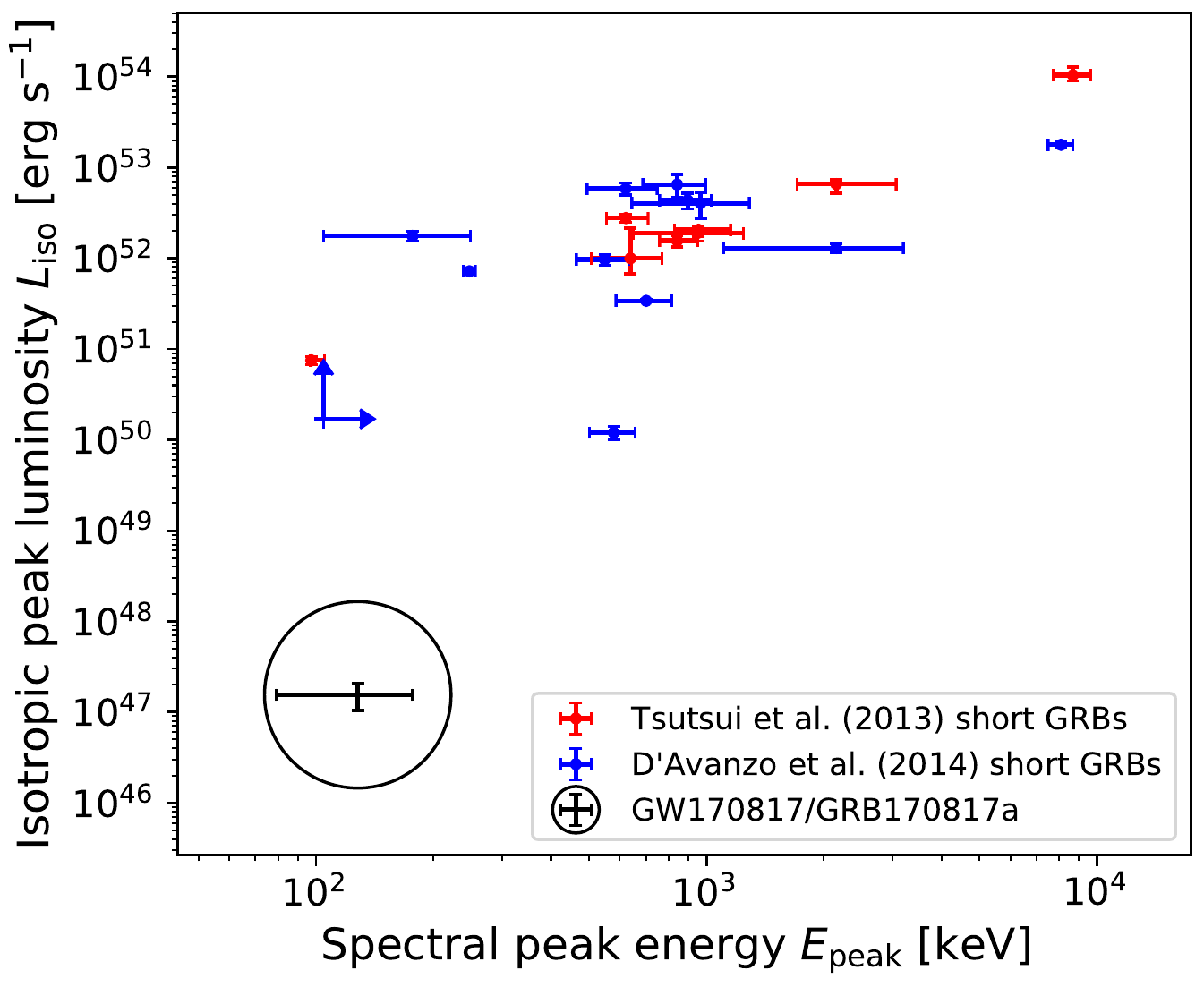}
}
\figcaption{ \textit{\textbf{Left}}: \Chandra\ X-ray flux light curve of GRB 170817A (black points), in comparison to predicted X-ray short GRB afterglow light curves from \citet{vaneerten11} for a range of jet axis angles from the line of sight (colored lines). The X-ray observations constrain the jet to be  $\gtrsim$23$^\circ$ off-axis, making GRB 170817A the first observed off-axis short GRB. The grey shaded region shows the time period during which \Chandra\ observations are not possible due to Sun constraints. \textit{\textbf{Right}}: $\gamma$-ray peak isotropic luminosity $L_\mathrm{iso}$ and $\gamma$-ray spectral peak energy $E_\mathrm{peak}$ of GRB 170817A (black), in comparison to samples of short GRBs from \citet[][red points]{tsutsui13} and \citet[][blue points]{davanzo2014}. The low luminosity of prompt $\gamma$-ray emission from GRB 170817A in comparison to other short GRBs may also support an off-axis jet interpretation.
}
\label{fig:lc}
\end{figure*}

For each of the two ObsIDs, we extracted spectra and response files for the X-ray sources using CIAO's {\tt specextract} tool. We then fit the spectra using XSPEC v12.9.0 \citep{Arnaud96}, with atomic cross sections and abundances from \citet{Verner96} and \citet{Wilms00}, respectively. The data from the two observations were first jointly-fit with an absorbed power law ($S\propto E^{-\Gamma}$). The absorption column in all cases was fixed to N\textsubscript{H}$=7.5 \times10^{20}$~cm$^{-2}$, derived by converting the Galactic optical extinction A$_{\rm V}=0.338$ \citep{Schlafly11} to a hydrogen column density via the relation N\textsubscript{H}(cm$^{-2})\approx2.21\times10^{21}$A$_{\rm V}$  \citep{Guver09}. The photon index was tied between the data sets while the normalization was left free. To obtain better constraints on both the photon indices and fluxes, the spectra (and response files) from the two observations were co-added and fit again with an absorbed power law. The resultant best-fit parameters are shown in Table \ref{tab:specvals} and Figures \ref{fig:fluxes} and \ref{fig:lc}.

The X-ray counterpart of the optical source SSS17a was well-fit with a power law index of $\Gamma\sim2.4$. Analysis of the spectra did not reveal a statistical difference in either the flux values or the count rates of SSS17a between the two observations. When we co-added the spectra and response files, the improved statistics yielded an absorbed flux of $\sim3.6\pm0.1\times 10^{-15}$~erg~s$^{-1}$~cm$^{-2}$. This is consistent with the upper limits observed by \Swift\ \citep[][see also Figure \ref{fig:fluxes}]{evans17}.  

The brightest of the four sources, the host galaxy NGC~4993, was well-fit with a power law index of $\Gamma=1.5\pm0.4$ and an absorbed 0.3-8~keV flux $F_{0.3-8~\mathrm{keV}}=1.3\pm0.2\times10^{-14}$~erg~s$^{-1}$~cm$^{-2}$. The soft energy excess visible in Figure \ref{fig:spec} may indicate the presence of thermal emission from a gaseous component in the galaxy. CXOU~J130948 has a similar photon index ($\Gamma=1.3\pm0.8$), though it has a lower flux of $F_{0.3-8~\mathrm{keV}}=4\pm0.1\times10^{-15}$~erg~s$^{-1}$~cm$^{-2}$. In contrast, CXOU~J130948 has a very hard spectrum with $\Gamma\approx-0.4\pm0.8$---this hard X-ray emission is also evident in Figure \ref{fig:image}, where the source appears visually blue in the three-color image. 

Assuming that these four sources are at the distance of the galaxy NGC 4993 \citep[$D_L = 42.5$ Mpc, ][]{daCosta98}, we derive the 0.3-10 keV X-ray luminosities listed in Table \ref{tab:specvals}. NGC 4993's X-ray luminosity from the combined, deep \Chandra\ observation is $L_{0.3-10~\mathrm{keV}} = 8.7^{+0.8}_{-0.9}\times10^{38}$~erg~s$^{-1}$, consistent with the X-ray luminosity of a lenticular E/S0-type galaxy \citep[e.g.,][]{kim15}. 

In addition to these two observations $\sim$2 weeks post-trigger, \citet{margutti21648} reported a non-detection of SSS17a from a $\sim$25~ks \Chandra\ observation 2 days after the detection of GRB 170817A.  Approximately 9 days post-trigger, \citet{troja21765} subsequently reported a \Chandra\ detection with a 50~ks exposure, though no flux values were reported.  The observations all share similar pointings and observing modes.  Based on the source (non-)detection status, we were able to use the response files, background spectra, and best-fit spectral parameters from the data reported here to simulate SSS17a emission from the other observations.  An upper limit of $F_{0.3-10~\mathrm{keV}}\approx3.4\times10^{-15}$~erg~s$^{-1}$~cm$^{-2}$ was obtained for the non-detection, and we place a lower limit of $F_{0.3-10~\mathrm{keV}}\approx2.8\times10^{-15}$~erg~s$^{-1}$~cm$^{-2}$ on the 9-day post-trigger detection (see Figs. \ref{fig:fluxes} and \ref{fig:lc}). These estimates assume no additional extenuating circumstances during the observation, such as strong solar background flares, bad pixels, etc.

\section{Discussion}
\label{sec:discussion}
\subsection{X-ray Evidence for an Off-Axis Short GRB}
It is challenging to explain our \Chandra\ X-ray detection of the afterglow of GRB 170817A at 16 days post-trigger, in combination with the \Chandra\ non-detection at 2 days post-trigger, and the detection at 9 days post-trigger, as described in Section \ref{sec:obssec}. For short GRBs, standard afterglow models predict that after the prompt emission fades on timescales of $<$2s post-burst, the relativistic jet will be decelerated by the ambient medium. This leads to X-ray emission that decays as $t^{-2}$ on the timescales of $10^{5-6}$s, i.e., exactly the timescales covered by our X-ray observations. Figure \ref{fig:lc} (left panel) displays our X-ray light curve of the afterglow of GRB 170817A, including the detection at 16 days post-trigger, an upper limit for the non-detection at 2 days post-trigger, and a lower limit for the detection at 9 days post-trigger. Figure \ref{fig:lc} (left panel) also displays theoretical 1.5 keV X-ray afterglow light curves for short GRBs for a range of jet axis angles, scaled to the observed flux of our \Chandra\ X-ray observations at 16 days. These light curve models are from the relativistic hydrodynamic simulations of \citet{vaneerten11}, which includes radiative transfer for synchrotron emission, and assumes that the beaming-corrected total energy in both jets is 10$^{48}$ erg, the number density of the ambient medium is 10$^{-3}$ cm$^{-3}$, and the jet half opening angle is 11$^\circ$. The model light curve in Figure \ref{fig:lc} for a jet observed at $0^\circ$ off-axis (i.e., directly along the line of sight) predicts afterglow X-ray emission that is a factor of $>$10$^3$ higher than the upper limit from the non-detection at 2 days post-trigger. Thus, the standard on-axis GRB scenario is disfavored by these X-ray observations.

GRB jets observed off-axis can produce afterglow emission that is faint at early times, but becomes luminous and observable as the jet beaming becomes less severe and the jet opening angle spreads into the line of sight \citep{granot02}. Figure \ref{fig:lc} (left panel) also displays model light curves at a range of jet axis angles from the line of sight. Our observed X-ray light curve is consistent with afterglow models for an off-axis short GRB, with jet axis angle of $\gtrsim$23$^\circ$. If confirmed, this makes GRB 170817A the first observed off-axis short GRB, in addition to the first electromagnetic counterpart to a GW event.

Further X-ray monitoring of GRB 170817A can tightly constrain the jet axis angle. If the jet axis angle is at $\gtrsim$23$^\circ$, the X-ray afterglow has already reached its peak and will continue to fade. However, at larger jet axis angles, the X-ray afterglow can still be brightening (e.g., see the model light curve for a 46$^\circ$ axis angle in Figure \ref{fig:lc}), until reaching a peak at up to 430 days post-burst for a jet at a 90$^\circ$ axis angle, using our assumed model parameters. Deep \Chandra\ observations after sunblock ($\sim$December 2017; Figure \ref{fig:lc}) could easily distinguish between these possibilities. 

The off-axis GRB scenario also predicts other multi-wavelength properties, including late-time radio emission from the afterglow that peaks on timescales of order 10 days after the X-ray peak. Indeed, a previously-undetected radio source associated with SSS17a was reported approximately 15 days post-burst \citep{mooley21814,corsi21815}. Thus, continued multi-wavelength monitoring of GRB 170817A will be key to unveiling its nature and understanding its properties.

\subsection{Gamma-ray Evidence for an Off-Axis Short GRB}

Our off-axis short GRB interpretation of GRB 170817A may also be supported by the low luminosity of its prompt $\gamma$-ray emission. The prompt emission from off-axis GRBs is likely to be under-luminous in comparison to on-axis GRBs, since it is strongly beamed. Figure \ref{fig:lc} (right panel) compares the $\gamma$-ray peak isotropic luminosity ($L_\mathrm{iso}$) and $\gamma$-ray rest-frame spectral peak energy ($E_\mathrm{peak})$ of GRB 170817A, to a sample of 8 short GRBs from \citet{tsutsui13} and a sample of 12 short GRBs from \citet{davanzo2014}. The $L_\mathrm{iso}$ and $E_\mathrm{peak}$ of GRB 170817A are based on $Fermi$ GBM observations over the 10-1000 keV range reported by \citet{goldstein21528}, which encompasses the 128 keV spectral peak. The prompt $\gamma$-ray emission of GRB 170817A appears to be strongly under-luminous in comparison to other short GRBs, although GRB 170817A is at much lower redshift ($z=0.009$) compared to these other samples ($z \sim 0.5$ to 1). The minimum luminosity of short GRBs is not well-constrained observationally, making it difficult to definitively distinguish off-axis GRBs from faint on-axis GRBs. However, the fact that the luminosity of GRB 170817A is a factor of $\sim$10$^3$ fainter than the next faintest short GRB, while having $E_\mathrm{peak}$ that is not unprecedented, suggests that GRB 170817A is not simply a faint on-axis short GRB, but is instead consistent with the interpretation that it is viewed off-axis.

\vspace{-1mm}
\acknowledgments
We dedicate this work to the memory of Neil Gehrels, one of the original PIs for our \Chandra\ proposal and an active participant in the early months of the program. Neil's stewardship of \Swift\ has influenced our entire community---this short $\gamma$-ray burst and its coincidence with a LIGO-Virgo GW source would have thrilled him. The authors owe a debt of gratitude to Belinda Wilkes and the \Chandra\ scheduling, data processing, and archive teams. Their incredibly fast work was essential to making these time-sensitive observations possible. We thank our anonymous referee for their timely review and useful comments. We thank Sean McWilliams for his useful input. This work was supported by Chandra Award Number GO7-18033X, issued by the Chandra X-ray Observatory, which is operated by the Smithsonian Astrophysical Observatory for and on behalf of the National Aeronautics Space Administration (NASA) under contract NAS8-03060. D.H. acknowledges support from the Canadian Institute for Advanced Research (CIFAR). M.N. and J.J.R. acknowledge funding from the McGill Trottier Chair in Astrophysics and Cosmology. D.H., M.N., and J.J.R. also acknowledge support from a Natural Sciences and Engineering Research Council of Canada (NSERC) Discovery grant and a Fonds de recherche du Qu\'{e}bec--Nature et Technologies (FRQNT) Nouveaux Chercheurs grant. P.A.E. acknowledges UKSA support. J.A.K. acknowledges NASA grant NAS5-00136.

\newpage
\bibliographystyle{apj}

\end{document}